\documentclass[
reprint,
 amsmath,amssymb,
 aps,
]{revtex4-2}

\usepackage{graphicx}
\usepackage{dcolumn}
\usepackage{bm}
\usepackage[utf8]{inputenc}
\usepackage{xcolor}
\usepackage{hyperref}

\begin{document}

\preprint{APS/123-QED}

\title{The influence of strong coupling between single-photon source and spectral filter on photon statistics}

\author{Ivan V. Panyukov and Evgeny S. Andrianov}
 \affiliation{Moscow Institute of Physics and Technology, 9 Institutskiy pereulok, Dolgoprudny 141700, Moscow region, Russia;}
 \affiliation{Dukhov Research Institute of Automatics (VNIIA), 22 Sushchevskaya, Moscow 127055, Russia;}

\date{\today}

\begin{abstract}
	One of the most common approaches for coupling optical single-photon sources and photonic integrated circuits is to use a cavity.
	The cavity acts as a spectral filter that distorts the light spectrum and changes its statistical properties.
	But in the general case one should take into account not only spectral filtering of light but also the spectral filter influence on the single-photon source dynamics.
	We build an effective analytical model for description of the cavity influence on the photon statistics of light emitted by the single-photon source as spectral filtering only.
	We show that this model correctly describes the photon statistics even in a strong-coupling regime between the single-photon source and the spectral filter.
	Our results can be useful for analytical modeling of photon statistics of quantum emitters strongly coupled to various electromagnetic interfaces. 
\end{abstract}

\maketitle

\section{INTRODUCTION}

Single-photon sources (SPSs) have a low second-order autocorrelation function $g^{(2)}(0)$, so they have a wide range of applications including quantum cryptography~\cite{hughes1995quantum} , quantum information transfer~\cite{gschrey2015highly}, quantum computing~\cite{cai2013experimental}, quantum metrology~\cite{von2019quantum}, and information processing~\cite{fortsch2013versatile}. 

As the scale of photonic quantum circuits increases, the importance of technological scalability and miniaturization becomes more apparent.
Integrated optics opens an obvious way to implement large-scale photonic quantum circuits with a large number of compactly packed elements~\cite{tanzilli2012genesis}. 
Several characteristics are important for integrated optics: achievable density of optical components, low losses during light propagation, speed of switching, and changes in the properties of individual elements for the possibility of external control. 
It is also desirable to integrate both light sources and detectors into an optical chip.
The main method for direct integration of solid-state quantum emitters with embedded nanophotonic structures is a coupling with a cavity. 
Cavities for the integration of SPSs provide advantages related to the efficiency and stability of the device and its simple operation.
It is possible to use different types of cavities to solve these problems, for example, ring cavities~\cite{faraon2013quantum}, disk cavities~\cite{radulaski2019nanodiamond}, photonic crystal cavities~\cite{faraon2012coupling}, and Fabry--Pero cavities~\cite{benedikter2017cavity}.

Cavities distort the spectrum of light emitted by SPSs~\cite{ming2012plasmon}, amplifying some components of the spectrum and suppressing others.
In general, the statistical properties of individual spectral components may differ from the statistical properties of the entire light containing all spectral components.
Thus, cavities that are used to interface SPSs with photonic circuits act as spectral filters, affecting the single-photon properties of the resulting emission.
In order to analyze the effect of the spectral filter on the statistical properties of light emitted by the SPS one can consider the spectral filtering of SPS emitted light neglecting the spectral filter influence on the SPS dynamics, as it was done in many works related to the spectral filtering of quantum light~\cite{gonzalez2013two, lopez2017photon, del2012theory,  lopez2022loss}.
However, the spectral filter can affect the SPS dynamics. 
This corresponds to the reflection of the SPS radiation from the spectral filter back to the SPS.
In order to take into account such influence, one can use a theory that describes
(1) the spectral filtering of light and the influence of the spectral filter on the SPS dynamics without any approximations,
(2) the spectral filtering of light and the influence of the spectral filter on the SPS dynamics via an effective model, in the frame of which one should describe the spectral filtering only, so one can apply the spectral filtering theories of quantum light~\cite{gonzalez2013two, lopez2017photon, del2012theory,  lopez2022loss}.
The last method is obviously computationally simpler, which makes them more preferable for analyzing real photonic devices. 
However, there is a question at which conditions it is valid.

In this paper, we consider the SPS coupled with a photonic circuit by the single-mode cavity, which acts as the Lorentz spectral filter, on the statistical properties of light emitted by the SPS.
We model the SPS as a two-level system (TLS).
We show that one should take into account not only spectral filtering of light but also the spectral filter influence on the single-photon source dynamics to correctly describe the second-order autocorrelation function of the spectrally filtered light.
We build an effective model which allows us to consider the spectral filtering only.
In the framework of this model the interaction of the TLS with the cavity is equivalent to the interaction of TLS with effective relaxation rates with a spectral filter which transforms the TLS emission spectrum only, but does not affect the TLS dynamics.
We compare analytical calculation of $g^{(2)}(0)$ via the effective model and numerical calculation via the full quantum-mechanical model that includes the spectral filter and the SPS.
We show that the effective model correctly describes $g^{(2)}(0)$ for almost all parameter values.
We show that minor discrepancies occur when (1) the emission spectrum of the system consists of two barely resolved peaks, and (2) the energy transfer rate from the single-photon source to the spectral filter exceeds the single-photon source own losses by an order of magnitude.
We show that one can associate these discrepancies with the cavity induced non-Markovian effects in the TLS dynamics. 

\section{CAVITY AND SPECTRAL FILTER EQUIVALENCE}

As a model of SPS coupled to a photonic circuit via a cavity, we consider an incoherently pumped TLS interacting with a harmonic oscillator.
This system is described by the Jaynes--Cummings Hamiltonian, which has the form
\begin{equation} \label{Hamiltonian full system}
\hat H = \hbar \omega_{\rm TLS} \hat \sigma^\dag \hat \sigma + \hbar \omega_{\rm Cav} \hat a^\dag \hat a + \hbar \Omega (\hat a^\dag \hat \sigma + \hat a \hat \sigma^\dag ),
\end{equation}
where $\hat \sigma$ is a lowering operator of the TLS, $\hat a$ is a lowering operator of the harmonic oscillator, $\omega_{\rm TLS}$ is a transition frequency of the TLS, $\omega_{\rm Cav}$ is a cavity eigenfrequency, $\Omega$ is a Rabi constant, which characterizes the interaction energy between the TLS and the cavity.
The Hilbert space of the system is defined by the states $\vert g,n \rangle$ and $\vert e,n \rangle$, where
$g$ and $e$ stand for ground and excited states of the TLS, and $n$ stands for the number of photons in the cavity. 

An emission from the TLS is transmitted to the photonic integrated circuit through the cavity; therefore, the second-order autocorrelation function for the system is equal to the second-order autocorrelation function of the cavity emission
\begin{equation} \label{g2cav}
g^{(2)}(0) =
\frac{\langle \hat a^\dag \hat a^\dag \hat a \hat a \rangle}{\langle \hat a^\dag \hat a \rangle^2}.
\end{equation}
In the general case, the value of $g^{(2)}(0)$ differs from the value of the second-order autocorrelation function of an isolated TLS single-photon emission
$g^{(2)}_{\rm TLS}(0) = \langle \hat \sigma^\dag \hat \sigma^\dag \hat \sigma \hat\sigma \rangle/
\langle \hat \sigma^\dag \hat \sigma \rangle^2=0$.
Now we express $g^{(2)}(0)$ through the TLS operators correlations. 
We obtain the Heisenberg-Langevin equation for the operator $\hat a$
\begin{equation}
\frac{d\hat a}{dt}=-(\gamma_a /2+i\omega_a)\hat a-i\Omega \hat \sigma + \hat \eta,
\end{equation}
where $\gamma_a$ is the cavity decay rate and $\hat \eta$ is an environment operator, which stands for the cavity decay.
The cavity decay rate and noise are connected via the fluctuation dissipation theorem~\cite{mandel1996optical}. 
We assume that the cavity mode contains no photons in the initial moment of time.
One can formally integrate the last equation and obtain
\begin{multline} \label{cavity_field}
\hat a(t)=\hat a(0)e^{-(\gamma_a /2+i\omega_a)t} \\
-i \Omega\int_0^t dt' \hat \sigma(t')e^{-(\gamma_a /2+i\omega_a)(t-t')} \\
+\int_0^t dt' \hat \eta(t')e^{-(\gamma_a /2+i\omega_a)(t-t')}.
\end{multline}
The first term on the right hand side plays the role of an initial condition and describes the vacuum fluctuations of the cavity electromagnetic field.
Substituting the resulting expression for $\hat a(t)$ into the expression for $g^{(2)}(0)$ leads to an expression
\begin{equation} \label{g2_exact}
g^{(2)}(0) = \frac{G^{(2)}(t,0)}{I^2(t)} \Biggr|_{t \to +\infty},
\end{equation}
where
\begin{multline}
G^{(2)}(t,0) = \int\limits_{0}^{t}dt_1 f^*(t-t_1) \int\limits_{0}^{t}dt_2 f^*(t-t_2) \\\int\limits_{0}^{t} dt_3 f(t-t_3) \int\limits_{0}^{t}dt_4 f(t-t_4) \\
\langle \mathcal{T}_\rightarrow [\hat \sigma^\dag(t_1)\hat \sigma^\dag(t_2)] \mathcal{T}_\leftarrow [\hat \sigma(t_3)\hat \sigma(t_4)]\rangle,
\end{multline}
\begin{equation}
I(t) = \int\limits_{0}^{t}dt_1 f^*(t-t_1) \int\limits_{0}^{t}dt_2 f(t-t_2) \langle\hat \sigma^\dag(t_1)\hat \sigma(t_2)\rangle.
\end{equation}
Here $f(t)=\gamma_a e^{-(\gamma_a/2+i\omega_a)t}/2$. 
The stationary regime corresponds to the limit $t \to +\infty$.
The symbols $\mathcal{T}_\rightarrow$ and $\mathcal{T}_\leftarrow$ stand for the time-ordering.
The presence of the time-ordering is a consequence of taking into account the vacuum fluctuations of the cavity electromagnetic field, that is, the first term on the right hand side in the Eq.~(\ref{cavity_field})~\cite{knoll1986quantum}.

Note that the expression~(\ref{g2_exact}) for $g^{(2)}(0)$ is identical to the expression for the second-order autocorrelation function of the TLS emission passed through a Lorentz spectral filter with a central frequency $\omega_a$ and a width $\gamma_a$~\cite{del2012theory}. 
The filter transfer function is $f(t)$ and the complex transmission function $F(\omega)$ is the Fourier transform of $f(t)$
\begin{equation}
F(\omega)=\int_{-\infty}^{+\infty}dtf(t)e^{i\omega t}=\frac{i\gamma_a/2}{\omega-\omega_a+i\gamma_a/2},
\end{equation}
where we defined $f(t)=0$ at $t<0$ due to the causality principle.
Thus, the single-mode cavity is a physical realization of the Lorentz spectral filter.

Nevertheless, it should be noted that the correlations of the TLS operators in the expression~(\ref{g2_exact}) are generally calculated taking into account the dynamics of the cavity, that is, the spectral filter.
Interaction with the spectral filter affects the TLS dynamics.
Physically, this corresponds to the reflection of the TLS emission from the spectral filter.
However, in some cases one can neglect the cavity influence when calculating the correlations of the TLS operators in the expression~(\ref{g2_exact}), as it was done in a lot of works~\cite{gonzalez2013two, lopez2017photon, del2012theory,  lopez2022loss}.

Next, we calculate $g^{(2)}(0)$ in three ways. 
Firstly, we provide rigorous calculation which fully takes into account the cavity influence on the TLS dynamics. 
Secondly, we will build an effective model within which the entire influence of the cavity can be reduced to the spectral filtering of the effective TLS emission only.
Third, we neglect the cavity influence on the TLS dynamics.
In the first case we  perform numerical calculation via the expression~(\ref{g2cav}).
In the second and in the third cases we use the expression~(\ref{g2_exact}) and obtain analytical expressions.
Next, we assume that the TLS and the cavity frequencies coincide $\omega_{\rm TLS}=\omega_{\rm Cav}=\omega_0$, that is, the spectral filter is set to the transition frequency of the TLS.

\section{DYNAMICS OF THE SYSTEM}

\subsection{Full quantum-mechanical model including the TLS and the spectral filter} \label{rig_dyn}
The current state of the TLS and the cavity is characterized by the density matrix of the system $\hat \rho$ that obeys the Lindblad equation
\begin{multline} \label{Lindblad full system}
\frac{d \hat{\rho}}{d t}
=
-\frac{i}{\hbar}\left[ \hat H, \hat \rho \right]
+
L_{{\rm diss}}\left[\hat{\rho}\right]
+
L_{{\rm deph}}\left[\hat{\rho}\right]
+
L_a\left[\hat{\rho}\right] \\
+
L_{{\rm pump}}\left[\hat{\rho}\right]
,
\end{multline}
where $L_{j}[\hat \rho]$ are the Lindblad superoperators that describe the relaxation dynamics of the system.
Consider each of the Lindblad superoperators from~(\ref{Lindblad full system}) in more detail.
Superoperator $L_{\rm diss}\left[\hat{\rho}\right]$ describes the decay of the TLS
\begin{equation} \label{L diss-TLS}
L_{\rm diss}\left[\hat{\rho}\right]
=
\frac{\gamma_{\rm diss}}{2} \left( 2\hat \sigma \hat \rho \hat \sigma^\dag - \hat \sigma^\dag \hat \sigma \hat \rho - \hat \rho \hat \sigma^\dag \hat \sigma \right), 
\end{equation}
where $\gamma_{\rm diss}$ is the decay rate of the TLS.
Superoperator $L_{\rm deph}\left[\hat{\rho}\right]$ describes the dephasing of the TLS 
\begin{equation} \label{L deph-TLS}
L_{\rm deph}\left[\hat{\rho}\right]
=
\frac{\gamma_{\rm deph}}{4} 
\left( 
\hat \sigma_z \hat \rho \hat \sigma_z - \hat \rho
\right),
\end{equation}
where $\gamma_{\rm deph}$ is the pure dephasing rate, and  $\hat \sigma_z=[\hat \sigma^\dag, \hat \sigma]$ is the population inversion operator.
The pure dephasing rate usually greatly exceeds other relaxation rates~\cite{albrecht2014narrow}.
In this paper, we will assume $\gamma_{\rm deph}=10^4\gamma_{\rm diss}$.
Superoperator $L_a\left[\hat{\rho}\right]$ describes the decay of the cavity
\begin{equation} \label{L diss-Cav}
L_a\left[\hat{\rho}\right]
=
\frac{\gamma_a}{2} \left( 2\hat a \hat \rho \hat a^\dag - \hat a^\dag \hat a \hat \rho - \hat \rho \hat a^\dag \hat a \right),
\end{equation}
where $\gamma_{a}$ is the decay rate of the cavity which is also its linewidth.
We also use the Lindlblad superoperators formalism to describe incoherent pumping of the system.
Superoperator $L_{\rm pump}\left[\hat{\rho}\right]$ describes the incoherent pumping of the TLS
\begin{equation} \label{L pump-TLS}
L_{\rm pump}\left[\hat{\rho}\right]
=
\frac{\gamma_{\rm pump}}{2} \left( 2\hat \sigma^\dag \hat \rho \hat \sigma - \hat \sigma \hat \sigma^\dag \hat \rho - \hat \rho \hat \sigma \hat \sigma^\dag \right), 
\end{equation}
where $\gamma_{\rm pump}$ is the pumping rate.
In the incoherent pumping scheme, one excites electrons to the levels of the SPS higher than the working level.
The electrons on these levels are sometimes called hot electrons.
The initial excitation then transits to the working level of the SPS~\cite{laucht2009dephasing}.
Eq.~(\ref{L pump-TLS}) describes this process as an effective transition from ground state to the working level of the  SPS.
This model is valid for quantum dots, single molecules, NV- and SiV- centers, two-dimensional materials.
In this paper, we will assume $\gamma_{\rm pump}=0.1\gamma_{\rm diss}$.
We consider the stationary problem, which stands for pumping long enough for all transients to end in the system.

The dipole moment relaxation rate of the isolated TLS is described by the parameter $\Gamma=(\gamma_{\rm diss}+\gamma_{\rm pump}+\gamma_{\rm deph})/2$.  The energy relaxation rate of the isolated TLS is described by the parameter $\gamma=\gamma_{\rm diss}+\gamma_{\rm pump}$.

\subsection{Effective model} \label{markov_dyn}

Due to the small pumping rate $\gamma_{\rm pump} \ll \gamma_{\rm diss}$, one can limit the Hilbert space of our system to three states: $\vert g,0 \rangle$, $\vert g,1 \rangle$, and $\vert e,0 \rangle$. 
Using Eq.~(\ref{Lindblad full system}), we obtain equations for the expectation values of the system operators.
We denote $a=\langle \hat a \rangle e^{i\omega_0 t}$ and $\sigma=\langle \hat \sigma \rangle e^{i\omega_0 t}$.
The equations for $a$ and $\sigma$ have the form
\begin{equation} \label{ep_1}
\frac{da}{dt}=-\frac{\gamma_a}{2} a - i \Omega \sigma,
\end{equation}
\begin{equation} \label{ep_2}
\frac{d\sigma}{dt}=-\Gamma \sigma - i \Omega a.
\end{equation}
One can distinguish a weak coupling regime $4\Omega<|2\Gamma-\gamma_a|$, and a strong coupling regime $4\Omega>|2\Gamma-\gamma_a|$.
In the weak coupling regime the system~(\ref{ep_1})--(\ref{ep_2}) eigenvalues are real.
This corresponds to one peak in the emission spectrum of the system.
In the strong coupling regime the system~(\ref{ep_1})--(\ref{ep_2}) eigenvalues have imaginary parts.
This corresponds to the appearance of two peaks in the emission spectrum of the system.
When these peaks are resolved, it stands for the coherent regime $\Omega>2\Gamma+\gamma_a$.
Using Eq.~(\ref{Lindblad full system}), we obtain equations for the energy variables
\begin{equation} \label{na_eq}
\frac{d n_a}{dt}=-\gamma_a n_a + \Omega J,
\end{equation}
\begin{equation} \label{nsigma_eq}
\frac{d n_{\sigma}}{dt}=
-\gamma_{\rm diss} n_{\sigma}
+\gamma_{\rm pump}(1-n_{\sigma})
-\Omega J,
\end{equation}
\begin{equation} \label{J_eq}
\frac{d J}{dt}=-\left(\frac{\gamma_a}{2} + \Gamma \right) J - 2 \Omega n_a + 2 \Omega n_{\sigma},
\end{equation}
where $n_a=\langle \hat a^\dagger \hat a\rangle$ is the cavity photon number, $n_{\sigma}=\langle \hat \sigma^\dagger \hat \sigma\rangle$ is the probability of the TLS to be in the excited state, and $J=\langle i(\hat a^\dagger \hat \sigma - \hat a \hat \sigma^\dagger) \rangle$ is the energy flow from the TLS to the cavity.

Consider Eq.~(\ref{nsigma_eq}).
The first term on the right hand side corresponds to the TLS own losses, the second term corresponds to the energy received by the TLS due to the incoherent pumping, and the third term corresponds to the TLS energy transfer to the cavity.
Thus, there are two channels of the TLS energy losses: the own losses and the energy transfer to the cavity.
The ratio of the TLS energy transfer rate to the cavity to the TLS own losses rate in the stationary regime is defined by the parameter $\beta=\Omega J/\gamma_{\rm diss} n_{\sigma}$.
Next, we also call the parameter $\beta$ as efficiency.

Consider an equation for $\sigma$.
One can integrate the equation~(\ref{ep_1}) with the initial condition $a(0)=0$ and substitute the result into the equation~(\ref{ep_2}) 
\begin{multline} \label{sigma_eff}
\frac{d\sigma}{dt}=-\Gamma \sigma - \Omega^2 \int_0^t dt'\sigma (t')e^{-\gamma_a(t-t')/2} \\
\approx -\left(\Gamma + \frac{2\Omega^2}{\gamma_a}\right)\sigma.
\end{multline}
The approximate equality stands for taking $\sigma(t')$ at $t'=t$ and the subsequent transition to the limit $t \to +\infty$.
Thus, the previous time moments do not contribute to the dynamics of the TLS at the current time moment, so one can consider this approximation as a Markovian.
Using Eq.~(\ref{na_eq})--(\ref{J_eq}), taking into account the initial conditions $a(0)=J(0)=0$, we obtain an equation containing only $n_{\sigma}$
\begin{multline} \label{spsm_int}
\frac{d n_{\sigma}}{dt}=
-2\Omega^2 K(t) n_{\sigma}(0)
-\gamma n_\sigma+\gamma_{\rm pump} \\
-
2\Omega^2(\gamma_a-\gamma)\int_0^t dt' K(t-t')n_{\sigma}(t') \\
+
2\Omega^2 \gamma_{\rm pump}\int_0^t dt' K(t-t'),
\end{multline}
where 
\begin{multline}
K(t)=e^{-(3\gamma_a+2\Gamma)t/4} \\
\frac{\sinh{\left(\sqrt{(\gamma_a/2-\Gamma)^2-16\Omega^2}t/2\right)}}{\sqrt{(\gamma_a/2-\Gamma)^2-16\Omega^2}/2}.
\end{multline}
We also make the approximation in the equation for $n_{\sigma}$ in the same way as it was made it the equation for $\sigma$
\begin{multline} \label{spsm_eff}
\frac{d n_{\sigma}}{dt}\approx
-\left(\gamma+\frac{4\Omega^2(\gamma_a-\gamma)}{\gamma_a(2\Gamma+\gamma_a)+8\Omega^2}\right) n_{\sigma} \\
+\gamma_{\rm pump}\frac{\gamma_a(2\Gamma+\gamma_a)+4\Omega^2}{\gamma_a(2\Gamma+\gamma_a)+8\Omega^2}.
\end{multline}
We have dropped the initial condition $n_{\rm\sigma}(0)$, since we consider the stationary dynamics, which is not affected by it.
Thus, our approximations lead to Eq.~(\ref{sigma_eff}) and~Eq.(\ref{spsm_eff}), which describe the effective isolated TLS with a Hamiltonian $\hat H^{\rm (eff)}=\hbar \omega_0 \hat \sigma^\dagger \hat \sigma$ and relaxation rates
\begin{equation} \label{diss_eff}
\gamma_{\rm diss}^{\rm (eff)}=\gamma_{\rm diss}+\frac{4\Omega^2(\gamma_a-\gamma_{\rm diss})}{\gamma_a(2\Gamma+\gamma_a)+8\Omega^2},
\end{equation}
\begin{equation} \label{pump_eff}
\gamma_{\rm pump}^{\rm (eff)}=\gamma_{\rm pump}\frac{\gamma_a(2\Gamma+\gamma_a)+4\Omega^2}{\gamma_a(2\Gamma+\gamma_a)+8\Omega^2},
\end{equation}
\begin{equation} \label{deph_eff}
\gamma_{\rm deph}^{\rm (eff)}=\gamma_{\rm deph}+
\frac{4\Omega^2(\gamma_a(2 \Gamma+\gamma)+8 \Omega^2)}{\gamma_a\left(\gamma_a(2\Gamma+\gamma_a)+8\Omega^2\right)}.
\end{equation}

\subsection{The spectral filter influence neglect on the TLS dynamics}

Another approximation is to neglect the influence of the spectral filter on the TLS dynamics. 
In such case our system also reduces to the isolated two-level system, but its the relaxation rates remain unchanged.
From mathematical point of view, it means that we should put $\Omega = 0$ when obtaining equations that describe the TLS dynamics via the Lindblad equation~(\ref{Lindblad full system}).
After that, one can use Eq.~(\ref{g2_exact}) and the quantum regression theorem~\cite{scully1997quantum, breuer2002theory} to calculate the second-order autocorrelation function of the filtered light when obtaining the TLS operators correlations (see Sections below).

\begin{figure*}[t!]
\begin{center} 
	\includegraphics[scale=0.4]{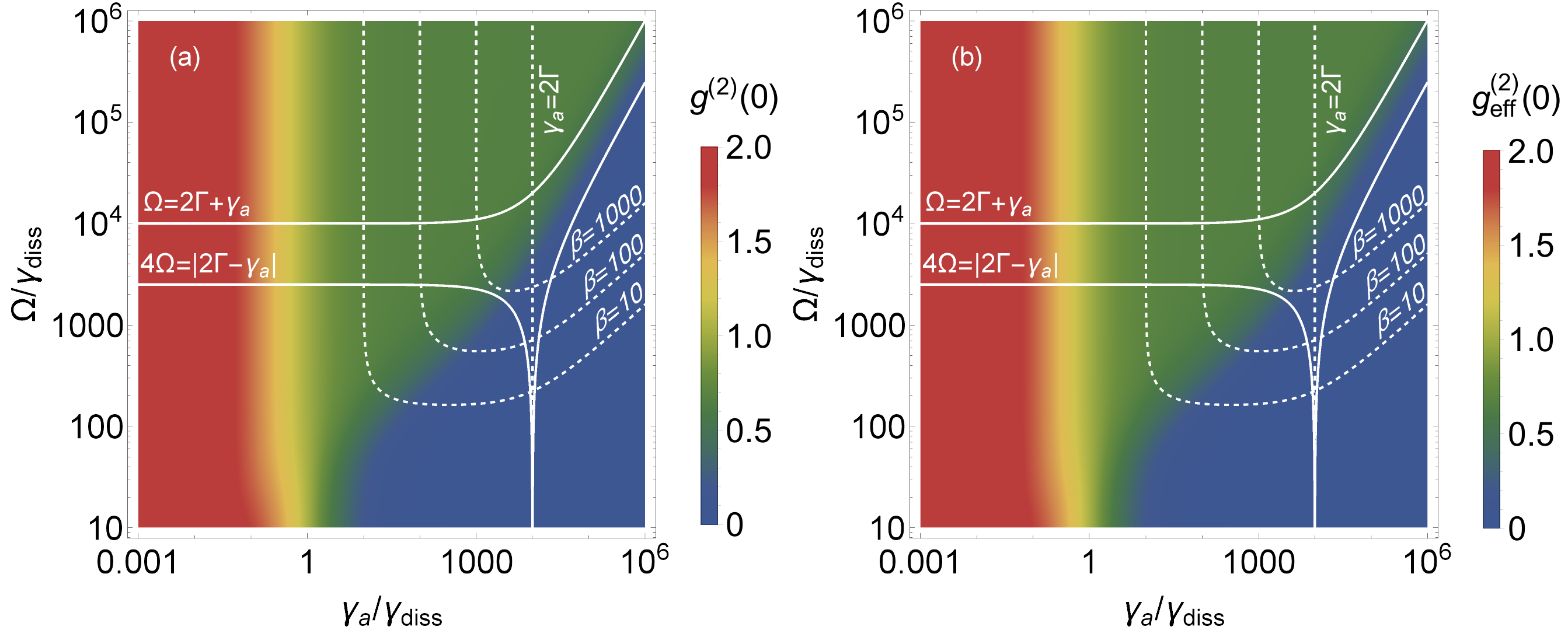} 
	\caption{ 
		Second-order autocorrelation function depending on the cavity linewidth $\gamma_a$ and the Rabi constant $\Omega$.
		(a) $g^{(2)}(0)$ calculated in the framework of the full quantum-mechanical model, including the TLS and the spectral filter.
		(b) $g^{(2)}_{\rm eff}(0)$ calculated via Eq.~(\ref{g2_markov}). 
	}  \label{dp1}
\end{center}
\end{figure*} 

\section{SECOND-ORDER AUTOCORRELATION FUNCTION}

\subsection{Full quantum-mechanical model including the TLS and the spectral filter}

In order to fully understand the influence of the spectral filter on the TLS dynamics, we perform the numerical calculation of $g^{(2)}(0)$ via the rigorous dynamics (Section~\ref{rig_dyn}) of the system.
We use the expression~(\ref{g2cav}) and the Lindblad equation for the density matrix~(\ref{Lindblad full system}).
One can see from a Fig.~\ref{dp1}(a) that reducing the cavity linewidth $\gamma_a$ leads to an increase in $g^{(2)}(0)$ from 0 to 2.
Note that the influence of the spectral filter on the TLS dynamics can be significant. 
For example, the value of the cavity linewidth at which $g^{(2)}(0)$ becomes far from zero increases as the Rabi constant $\Omega$ increases.
Note that it is possible to simultaneously achieve high efficiency $\beta$ and small $g^{(2)}(0)$ in the weak coupling regime $4\Omega<|2\Gamma-\gamma_a|$ and when $\gamma_a>2\Gamma \approx \gamma_{\rm deph}$.

\subsection{Effective model}

\begin{figure}[t!]
\begin{center} 
	\includegraphics[scale=0.37]{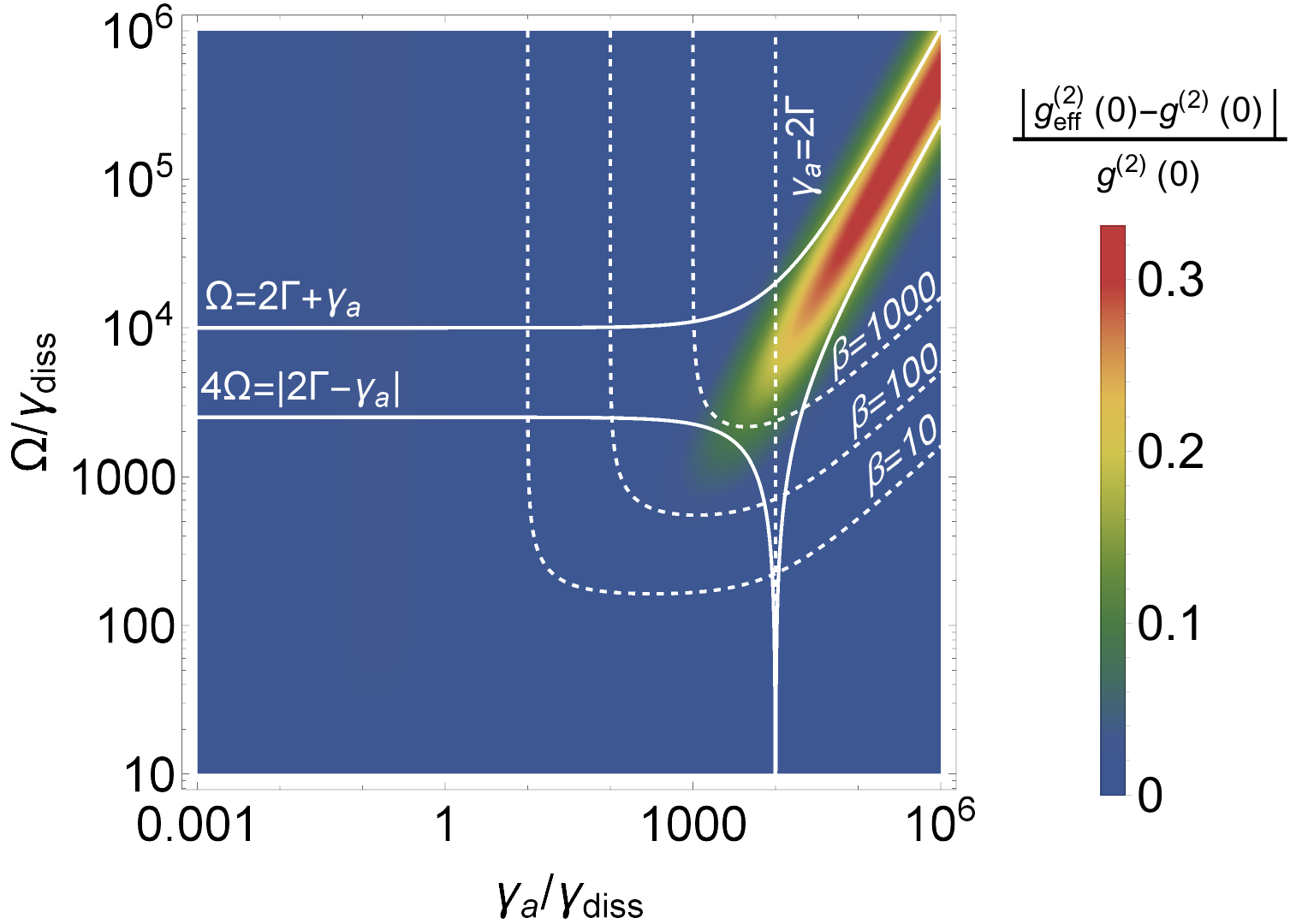}
\end{center}
\caption{ 
	Relative difference between $g^{(2)}(0)$ and $g^{(2)}_{\rm eff}(0)$.
}  \label{dp2} 
\end{figure} 

\begin{figure*}[t!]
\begin{center} 
	\includegraphics[scale=0.34]{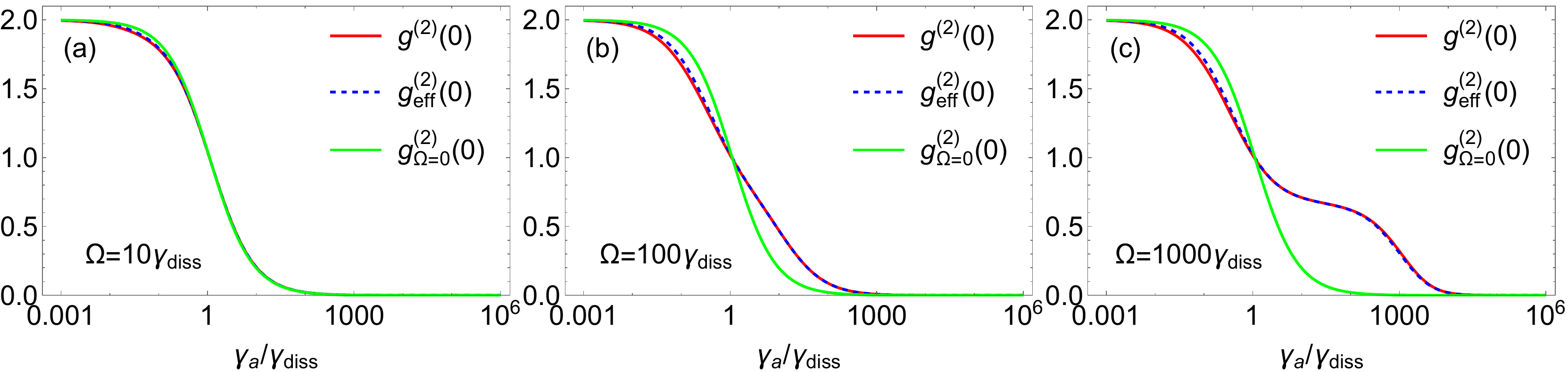}
\end{center}
\caption{ 
	Second-order autocorrelation function calculated rigorously, neglecting the spectral filter influence, and via the effective model, at different values of $\Omega$.}
\label{plot}
\end{figure*} 

Now we take the cavity influence on the TLS dynamics via the effective model.
As we have shown in Section~\ref{markov_dyn}, in such case our system reduces to an effective isolated TLS with modified relaxation rates. 
We use the expression~(\ref{g2_exact}) to calculate $g^{(2)}(0)$.
In order to calculate $g^{(2)}(0)$ one need to obtain correlations $\langle \mathcal{T}_\rightarrow [\hat \sigma^\dag(t_1)\hat \sigma^\dag(t_2)] \mathcal{T}_\leftarrow [\hat \sigma(t_3)\hat \sigma(t_4)]\rangle$ and $\langle\hat \sigma^\dag(t_1)\hat \sigma(t_2)\rangle$.
To do this, one should apply the quantum regression theorem, Eq.(\ref{sigma_eff}), and Eq. (\ref{spsm_eff}).
We obtain an expression
\begin{equation} \label{g2_markov}
g^{(2)}_{\rm eff}(0)=\frac{2\gamma^{\rm (eff)} (2\Gamma^{\rm (eff)}+\gamma_a)}{(\gamma^{\rm (eff)}+\gamma_a)(2\Gamma^{\rm (eff)}+3\gamma_a)},
\end{equation}
where $\Gamma^{\rm (eff)}=(\gamma_{\rm diss}^{\rm (eff)}+\gamma_{\rm pump}^{\rm (eff)}+\gamma_{\rm deph}^{\rm (eff)})/2$ 
and $\gamma^{\rm (eff)}=\gamma_{\rm diss}^{\rm (eff)}+\gamma_{\rm pump}^{\rm (eff)}$.

It can be seen from the Fig.~\ref{dp1} and the Fig.~\ref{dp2} that $g^{(2)}(0)$ and $g^{(2)}_{\rm eff}(0)$ are in a good agreement.
At sufficiently small $\beta$ the agreement between $g^{(2)}(0)$ and $g^{(2)}_{\rm eff}(0)$ takes place for all possible parameter values.
$g^{(2)}(0)$ and $g^{(2)}_{\rm eff}(0)$ do not differ in the bad cavity limit that corresponds to a region where 
(1) the weak coupling regime $4\Omega<|2\Gamma-\gamma_a|$, and 
(2) the large cavity decay rate $\gamma_a>2\Gamma \approx \gamma_{\rm deph}$ take place.
One can see that in this limit $\sigma(t')$ in Eq.~(\ref{sigma_eff}) and $K(t-t')$ in Eq.~(\ref{spsm_int}) have the crucial contribution at $t'\approx t$.
The contributions of the previous moments of time $t'<t$ to the TLS dynamics at the current time moment $t$ are exponentially suppressed due to the strong cavity decay.
The effective model also works good in the coherent regime $\Omega>2\Gamma+\gamma_a$, where two peaks in the emission spectrum of the system are clearly distinguishable.
In this limit $\sigma(t')$ in Eq.~(\ref{sigma_eff}) and $K(t-t')$ in Eq.~(\ref{spsm_int}) also have the crucial contribution at $t'\approx t$.
Contributions of the previous moments of time $t'<t$ to the TLS dynamics at the current time moment $t$ are averaged due to oscillations of $\sigma(t')$ and $K(t-t')$.

When $\beta$ reaches values of the order of $1000$, there is small difference between $g^{(2)}(0)$ and $g^{(2)}_{\rm eff}(0)$, when (1) the strong coupling regime $4\Omega>|2\Gamma-\gamma_a|$, and (2) the incoherent regime $\Omega<2\Gamma+\gamma_a$ take place.
Thus, the maximal difference between $g^{(2)}(0)$ and $g^{(2)}_{\rm eff}(0)$ occurs when the two following factors take place: 
(1) the TLS energy transfer rate to the cavity exceeds the TLS own losses by more than several orders of magnitude, 
(2) the emission spectrum of the system consists of two barely resolved peaks.
In this region previous moments of time $t'<t$ contribute to the TLS dynamics at the current time moment $t$.
Consequently, one can associate this region with the cavity-induced non-Markovian dynamics of the TLS.

\subsection{The spectral filter influence neglect on the TLS dynamics}

Now we neglect the influence of the spectral filter on the TLS dynamics. 
In such case we use the expression~(\ref{g2_exact}) to calculate $g^{(2)}(0)$.
In order to calculate $g^{(2)}(0)$ one also need to obtain correlations $\langle \mathcal{T}_\rightarrow [\hat \sigma^\dag(t_1)\hat \sigma^\dag(t_2)] \mathcal{T}_\leftarrow [\hat \sigma(t_3)\hat \sigma(t_4)]\rangle$ and $\langle\hat \sigma^\dag(t_1)\hat \sigma(t_2)\rangle$.
To do this, one should apply the quantum regression theorem and put $\Omega = 0$.
We obtain an expression
\begin{equation} \label{g2_neglect}
g^{(2)}_{\Omega=0}(0)=\frac{2\gamma (2\Gamma+\gamma_a)}{(\gamma+\gamma_a)(2\Gamma+3\gamma_a)}.
\end{equation}
This result is consistent with the calculation of the second-order autocorrelation function of light from an incoherently pumped TLS passed through a Lorentz spectral filter~\cite{lopez2022loss}.
As one can see from a Fig.~\ref{plot}, 
one can apply this result to calculate $g^{(2)}_{\Omega=0}(0)$ at $\Omega < 100\gamma_{\rm diss}$.
When $\Omega > 100\gamma_{\rm diss}$, one must not neglect the spectral filter influence on the TLS dynamics.
Moreover, one can not use this approach to describe the dependence on $\Omega$ of the cavity linewidth, at which $g^{(2)}(0)$ becomes far from zero.

\section{CONCLUSION}

We considered the change in the second-order autocorrelation function $g^{(2)}(0)$ of light from the incoherently pumped SPS when coupled to a photonic quantum circuit using the single-mode cavity.
We modeled the SPS as the TLS.
The interaction of the TLS with the cavity leads to a change in the system emission spectrum, which causes a change in the second-order autocorrelation function $g^{(2)}(0)$.
The influence of the cavity on $g^{(2)}(0)$ is equivalent to the effect of a Lorentz spectral filter having central frequency equal to the cavity frequency and having a width equal to the cavity linewidth.
As the the spectral filter width decreases, $g^{(2)}(0)$ increases from 0 to 2.

In general case, the spectral filter affects the dynamics of the TLS, and we show that one should take into account not only spectral filtering of light but such influence to.
We build the effective analytical model to describe photon statistics in the system.
In the framework of this model, the cavity influence on the second-order autocorrelation function can be reduced to spectral filtering only. 
Within this model the light is emitted by the single TLS with effective relaxation rates.
We showed that the effective model correctly describes the second-order autocorrelation function even in the strong coupling regime and provides analytical expressions.
Therefore, we showed that the interaction with the spectral filter only quantitatively, but not qualitatively, affects the TLS dynamics.
We shown that, when calculating the second-order autocorrelation function, the effective model is valid for all parameter values, as long as the energy transfer rate from the TLS to the cavity exceeds the TLS own losses by no more than several orders of magnitude.
In the opposite case, it may give incorrect results, if (1) strong coupling regime, and (2) incoherent regime take place.
We have also shown that one can associate this region with the cavity-induced non-Markovian effects in the TLS dynamics.

\section*{Acknowledgments}
I.V.P. and E.S.A. thank the foundation for the advancement of theoretical physics and mathematics “Basis”.

\bibliography{SPS_in_cavity}
\end{document}